\documentstyle[12pt,titlepage]{article}

\newtheorem{theorem}{Theorem}
\def\Re{\hbox{\rm I\kern-.18em R}}

\begin{document}

\title{\bf Nonintegrability and Chaos in  the Anisotropic Manev Problem}
\author{{\bf Florin Diacu\thanks{e-mail: diacu@math.uvic.ca, phone:
(250)721-6330, 
fax: (250)721-8962} {\rm and} 
Manuele Santoprete}\thanks{e-mail: msantopr@math.uvic.ca}\smallskip 
\\{Department of Mathematics and Statistics}\\
{University of Victoria, P.O. Box 3045}\\ {Victoria B.C., Canada,
V8W 3P4}}
\maketitle
\def \r#1{$^{#1}$}

\begin {abstract}
\def \Ref{\pn{\bf References}\bs\pn\parskip=5 pt\parindent=0 pt}
 
\noindent The anisotropic Manev problem, which lies at the intersection of
classical, quantum, and relativity physics, describes the motion 
of two point masses in an anisotropic space under the influence of a Newtonian 
force-law with a relativistic correction term. Using an extension of the 
Poincar\'e-Melnikov method, we first prove that for weak anisotropy, chaos 
shows up on the zero-energy manifold. Then we put into the evidence a class 
of isolated periodic orbits and show that the system is nonintegrable. 
Finally, using the geodesic deviation approach, we prove the existence of a 
large non-chaotic set of uniformly bounded and collisionless solutions.

\vskip0.6cm

\noindent {\bf PACS}(2000): 05.45.Ac, 45.10.Hj, 45.50.Jf

\medskip

\noindent {\bf Key words}:  Manev problem, anisotropy, chaos, geodesic
deviation

\end{abstract}
\def \bs{\bigskip}

\bs\bs
\def \pn{\par\noindent}
\quad

\def\R{{\bf R}}
\def \d{{\rm d}}

\def \nhy {nonhyperbolic}
\def \pd{\partial}
\def \pn{\par\noindent}

\def \en{\eqno}

\def\.#1{\dot #1}

~ \vskip 1 truecm

\parskip=0pt

\def\ref#1{$^{#1}$}

\def \de{\delta}
\def \th{\theta}
\def \Th{\Theta}
\def \ep{\epsilon}
\def \a {\alpha}
\def \ga{\delta}
\def \be{\beta}
\def \b{\beta}
\def \W{{\cal W}}
\def \r#1{$^{#1}$}
\def \( {\big( }
\def \) {\big) }
\def \intR {\int_{-\infty}^{+\infty}}
\def \grad {\nabla}
\def \ov{\over}
\def \G {{\pd W\big(R(t),t+t_0\big)\ov{\pd r}}}
\def \Gs{{\pd W\big(R(s),s+t_0\big)\ov{\pd r}}}
\def \q {\quad}
\def \qq {\qquad}
\def \bar{\overline}
\def \1{{\bf I}}
\def \2{{\bf II}}
\newtheorem{proposition}{Proposition}
\def \Ref{\pn{\bf References}\bs\pn\parskip=5 pt\parindent=0 pt}

\def\.#1{\dot #1}

\section*{\large\bf I. Introduction}

\noindent The Manev problem studies the dynamics of two point masses in a flat
space for an attraction law given by a potential of the form ${A\over r}+
{B\over r^2}$, where $r$ is the distance between particles and $A,B>0$ 
are constants. The correction term to the Newtonian potential 
provides a classical approximation to general relativity \cite{1}, \cite{2}.
The anisotropic Manev 
problem, proposed by the first author in the mid 1990s [3], replaces the
flat space 
with an anisotropic one.   

The main reason for considering this question was that of seeking
similarities between 
classical mechanics, quantum theory, and general relativity. Its study was
inspired by 
the anisotropic Kepler problem introduced by Gutzwiller in the early 1970s
\cite{4}, \cite{5}. 
Gutzwiller aimed to find connections between classical and quantum
mechanics. His
interest in the anisotropic Kepler problem was aroused by an old quantum
mechanical
question related to a paper of Einstein \cite{6}. 

Even though for a special class of (integrable) systems the
Born-Sommer\-feld-Einstein 
conditions \cite{6} seemed appropriate for describing a Coulomb limit of
quantum theory, 
it was unclear how to find a classical approximation for ergodic systems. 
Two qualities made the anisotropic Kepler problem suitable for addressing
this issue:
its chaotic character and its suitability to model various physical
phenomena, as for
example those encountered in certain semiconductors \cite{4}. Using a
classical evaluation 
of the quantum Green's formula, Gutzwiller \cite{7} found an approximate
spectrum, in good 
agreement with previous quantum calculations. All these properties of the
anisotropic Kepler 
problem raised hopes for the study of the Manev analogue, which also brought
general 
relativity into the game.

Some of the qualitative features of the anisotropic Manev problem problem
have already 
been studied. In a previous paper we proved the existence of an open,
connected invariant 
manifold of uniformly bounded, collisionless orbits that lie within the
negative-energy 
manifold \cite{8}. In \cite{3} it was shown that the differential equations
describing 
the anisotropic Manev problem exhibit properties characteristic to all three
aforementioned 
branches of physics. That paper also brought arguments favoring the
nonintegrability of the 
system by putting into the evidence a network of heteroclinic orbits within
the zero-energy 
manifold. Up to now, however, the existence of chaos and the
nonintegrability of the system 
have not been rigorously proved.

The first result of this paper shows the presence of chaos for a small set
of solutions 
within the zero-energy manifold (see Section~III). We will base our proof on
an extension 
of the Poincar\'e-Melnikov method developed in an earlier paper \cite{9}.
This method was 
used there to show the presence of chaos in several problems, including the
restricted 
circular 3-body problem and the Gyld\'en problem. The latter had been
earlier identified as 
a degenerate case for applying the Poincar\'e-Melnikov method \cite{10}. In
the 3-body problem 
the perturbation function, $W=W(r,\th, t)$, where $(r,\th)$ are the polar
coordinates, 
appeared in its most general form. In the Gyld\'en problem this function was
isotropic, 
$W=W(r,t)$. From the methodological point of view, we will now complete the
applicative 
picture by offering a time-independent but anisotropic perturbation function
of the form 
$W=W(r,\th)$, which is specific to the anisotropic Manev problem (see
Section~II).

It is interesting to remark that different perturbations break up different
symmetries, 
which affect the Melnikov integrals. In the general case both the
homogeneity of 
time and the rotational symmetry are destroyed. This leaves us with two
independent 
Melnikov conditions. In the Gyld\'en problem the rotational symmetry is
preserved (the 
angular momentum is constant), which leads us to one Melnikov condition. In
the present 
case only the time homogeneity is preserved (through the conservation of the
Hamiltonian), 
so the only object of study is the other Melnikov condition (see Section~III).

Our second result proves that the anisotropic Manev problem has no other
analytic integrals 
beyond the Hamiltonian function. To show this, we use a method designed by
Poincar\'e 
\cite{11}, \cite{12}, which we present in Section~IV. The idea is to prove
the existence of 
at least one isolated periodic orbit, which obstructs the occurrence of
other analytic 
integrals. To apply Poincar\'e's method we need to write the perturbation as
a Fourier 
series. This leads to some technical difficulties, which we overcome in
Section~V. In 
Section~VI we prove the existence of many isolated periodic orbits in the
hypothesis of 
weak anisotropy. We then show that this implies nonintegrability. The
existence of periodic 
orbits, however, is an interesting result in itself, which raises hopes for
applying 
Gutzwiller's formula to this problem in the context of semiclassical studies
of the atom. 

In the last part of the paper we are concerned with the qualitative behavior of
collisionless orbits for which the two particles stay close to each other. In
Section~VII we present the geodesic deviation method \cite{13}, which
provides a 
criterion for excluding the presence of chaos. The idea is to transform the
equations
of motion into some equivalent ones. The new equations put the solutions of
the given 
system in a one-to-one correspondence with the geodesics of a certain
Riemannian 
manifold whose metric is determined by the initial equations of motion.
Measuring the 
deviation of the geodesics, we can draw conclusions about the behavior of
the initial 
solutions. If the geodesics diverge, chaos may appear; if they converge,
chaos is ruled 
out. In Section~VIII we show that there exists a positive-measure invariant
manifold of 
orbits that fail to encounter collisions but for which the particles stay
close to 
each other. Using the geodesic deviation method we then prove that the
corresponding 
geodesics converge, thus showing that the orbits of this manifold are not
chaotic.

\section*{\large\bf II. Equations of motion}

\noindent The (planar) anisotropic Manev problem is described by the
Hamiltonian 
$${\cal H}({\bf q,p})={1\over 2}{\bf p}^2-{1\over \sqrt{\mu
q_1^2+q_2^2}}-{b\over{\mu
q_1^2+q_2^2}}, \en(1)$$
where $b>0$ and $\mu>1$ are constants, ${\bf q}=(q_1,q_2)$ is the position
of one body with 
respect to the other considered fixed at the origin of the coordinate
system, and 
${\bf p}=(p_1,p_2)={\dot{\bf q}}$ is the momentum of the moving particle. 
The constant $\mu$ measures the strength of the anisotropy; the larger
$\mu$, the higher
the difference between the weakest and the strongest directions of
attraction. For $\mu=1$ 
we recover the classical (isotropic) Manev problem of which we have a
complete qualitative 
understanding \cite{14}, \cite{1}. Therefore, if $\mu$ is only slightly
larger than 1, we
are dealing with a {\it weakly anisotropic} case. Unlike in the isotropic
problem, the angular 
momentum $K(t)={\bf p}(t)\times{\bf q}(t)$ of the anisotropic case is not a
first integral 
of the system. This is because the rotational invariance of the Hamiltonian
breaks for 
$\mu>1$. Therefore we expect to encounter richer dynamics in the anisotropic
case.

In the first part of this paper we will be interested in aspects related to
weak 
anisotropy, i.e., in values of $\mu>1$ that are close to 1. To put into the
evidence 
the perturbative character of the anisotropy with respect to the isotropic
problem, 
we introduce the notations $\ep = \mu -1$ and $r=\sqrt{q_1^2+q_2^2}$. For
$\mu>1$ 
close to 1, the quantity $\ep>0$ is small, so the Hamiltonian (1) can be
viewed as
describing a perturbation of the Manev problem by writing 
$${\cal H}({\bf q,p})={1\over 2}{\bf p}^2-{1\over r}-{b\over r^2}+\ep
\Bigg({1 \over 2r}
+{b\over r^2}\Bigg) \cos^2 \th\equiv {\cal H}_0+\ep {\cal W}(r,\th).\en(2)$$

As in \cite{10}, we now consider the parabolic solutions of the unperturbed
problem 
(defined by the Hamiltonian ${\cal H}_0$) that lie on the zero-energy
manifold. These 
orbits, which play the role of homoclinic solutions corresponding to the
critical point 
at infinity ($r=\infty,\ \.r=0$), satisfy the equations 
$$\.r=\pm {\sqrt{2r-(k^2-2b)}\over r},\ \ \.\th={k\over{r^2}},\en(3)$$
where $k\not=0$ is the constant angular momentum of the unperturbed problem;
the 
negative (positive) sign corresponds to $t<0$ ($t>0$). From (3) we get that
$$\cases{\pm t={k^2-2b+r\over 3}\sqrt{2r-(k^2-2b)} + {\rm constant}\cr
\th=\pm 2\arctan{\sqrt{2r-(k^2-2b)}\over \sqrt{k^2-2b}}+ {\rm
constant}.\cr}\en(4)$$
Let  
$$R=R(t) \qq {\rm and}\qq \Th=\Th(t)\en(5)$$
be the expressions of $r$ and $\th$ with respect to $t$, obtained by
inverting equations 
(4) and assuming that $R(0)=r_{min} =k^2/2$ and $\Th(0)=\pi$. We will not
need their expressions;
as in \cite{9}, we only retain the information that $R$ is even and that
$\Th$ is odd.

\section*{\large\bf III. The Melnikov integral}

We now consider the problem defined by the Hamiltonian ${\cal H}$ given in
(2). We will call 
{\it homoclinic manifold} the set of solutions of the unperturbed system
that are 
doubly asymptotic to the point $r=\infty,~ \.r=0$. For $k \neq 0$, the
two-dimensional 
homoclinic manifold is described by $r=R(t-t_0)$ and $\vartheta = \Th (t-t_0) 
+ \th_0$, with arbitrary constants $t_0$ and $\th_0$.
     
It is interesting to remark that in this problem we are dealing with
negatively and 
positively asymptotic sets to the critical point $r=\infty,~ \.r=0$ and that
this
point would correspond to the ``critical point at infinity,'' obtained with
the help of McGehee's transformation \cite{15}
$$r={1 \over x^2},\qq \.r=y,\qq dt={1 \over x^3}ds.\en(6)$$    
But instead of using McGehee's technique, we will apply here the method
introduced in 
\cite{9}. It was shown there that under natural assumptions imposed on the 
perturbation function ${\cal W}(r,\th)$, one can guarantee the existence of
smooth solutions 
approaching the point $r=\infty,~ \.r=0$ for $t\rightarrow \pm \infty$. It
was also proved 
that the Poincar\'e map leads to infinitely many intersections of the stable
and 
unstable manifold for the corresponding fixed point.

In our case the perturbation arising from a weak anisotropy vanishes as $r
\rightarrow 
\infty$. This happens because ${\cal W}(r,\th) \sim 1/r$, which is exactly
condition (18) 
in \cite{9}. So we can write the Melnikov condition as in [9], with the only
amendment 
of dropping the time dependence. Therefore the two Melnikov integrals become 
$$M_1(\th_0)=\intR \!\!
\Bigg[\.R(t) {\pd {\cal W}\big(R(t),\Th(t)+\th_0\big)\ov{\pd r}} + 
\.\Th(t) {\pd {\cal W}\big(R(t),\Th(t)+\th_0\big)\ov{\pd\th}} \Bigg]\d t 
=0\en(7)$$
and
$$M_2(\th_0)=
\intR {\pd {\cal W}\big(R(t),\Th(t)+\th_0\big)\ov{\pd \th}}=0. \en(8)$$
Since $W$ vanishes as $t \rightarrow \pm \infty$, the first Melnikov
condition takes the form
$$M_1(\th_0)=\intR {\pd {\cal W}\big(R(t),\Th(t)+\th_0\big)\ov{\pd t}} 
\d t \equiv 0.\en(9)$$
$M_1$ is identically zero since the perturbation function $\cal W$ is
independent of time.
This simplifies our discussion because, unlike in the general case, we only
need to find 
the zeroes of the integral in (8). 

It is significant to remark, and easy to check, that the previous conditions can
also be written in terms of the first integrals of the unperturbed problem as
$$M_1(\th_0) = \intR \{{\cal H}_0,{\cal W}\}\big(\ldots)\ \d t = 0\en(10)$$
and
$$M_2(\th_0) = \intR \{K,{\cal W}\}\big(\ldots)\ \d t = 0.\en(11)$$
Notice that $M_1(\th_0)$ is identically zero. This resembles
a results obtained for the  Gyld\'en problem \cite{9}, \cite{10} and is
related to
the symmetries of the problem. In the Gyld\'en problem the perturbation is
independent
of the angle $\th$ but depends on time. This means that the perturbation
alters the 
homogeneity of time, so the Hamiltonian is not an integral of motion
anymore,  but it 
leaves the rotational invariance intact; thus the angular momentum is still
conserved. 
Therefore there is only one condition here, given by relation (9). On the
other hand 
in the anisotropic Manev problem the anisotropy of the space alters only the
rotational 
symmetry but not the homogeneity of time. This also leads to only one
condition, given 
by relation (8).

For the anisotropic Manev problem the Melnikov condition $M_2$ takes the form
$$M_2(\th_0) = \intR \sin[ 2( \Th(t)+\th_0)]\Bigg({1 \over R(t)}+{b \over
R(t)^2}\Bigg)d t = 0.\en(12)$$
Using some trigonometry the integral can be written as
$$M_2(\th_0)=  I_1 \cos 2 \th_0  +I_2  \sin 2\th_0,  \en(13)$$
where $I_1$ and $I_2$ are defined by the relations
$$\cases{I_1= \intR ({1 \over R(t)}+{b \over R(t)^2}) \sin 2 \Th(t) dt\cr
I_2= \intR ({1 \over R(t)}+{b \over R(t)^2}) \cos 2 \Th(t) dt.\cr}\en(14)$$
Since $R$ and $\Th$ are even and odd functions of time, respectively, the
integrand of 
$I_1$ is an odd function. Therefore $I_1 \equiv 0$, and finally $M_2$ can be
rewritten as
$$ M_2(\th_0)= I_2 \sin 2\th_0. \en(15)$$ 
Since we found simple zeroes of the Melnikov function, the following result
holds.

\begin{theorem} In some invariant set contained in the zero-energy manifold,
the 
weakly anisotropic Manev problem exhibits chaotic dynamics. 
\end{theorem}

This type of chaotic behavior, which usually takes place for a small set of
orbits, is 
induced by the infinitely many intersections of the stable and unstable
manifolds for 
the Poincar\'e map associated to the critical point at infinity. Let us
notice that the 
Poincar\'e-Birkhoff-Smale theorem \cite{16} does not directly apply to this
situation, which 
is degenerate. However, the existence of Smale horseshoes and of positive
topological 
entropy is not restricted to hyperbolic equilibria. This phenomenon is also
encountered 
in nonhyperbolic cases, as for example those arising when dealing with
area-preserving 
perturbations \cite{17}. Thus, Theorem~1 adds to the class of results
describing chaotic 
dynamics near degenerate equilibria.

\section*{\large\bf IV. Poincar\'e's method}

In this section we will present a classical result of Poincar\'e \cite{11},
\cite{12} 
in connection with the nonexistence of additional analytic integrals in
Hamiltonian 
systems. Based on the investigation of long periodic solutions, this
criterion is 
suitable for proving the nonintegrability of the anisotropic Manev problem
for weak 
anisotropy, as we will see in Section~V. 

First we need to write the unperturbed system, i.e. the (isotropic) Manev
problem, in 
terms of action-angle variables \cite{18}. The action variables are given by
$$\cases{I={{1 \over 2 \pi} \oint p_r dr= - \sqrt{ K^2-2b} +{1 \over 2}{\sqrt {2
\over |h|}}}\cr
K=q_1p_2-q_2p_1,\cr}$$
where $h$ is the energy constant and $K$ is the angular momentum. These
variables are 
defined for $h<0$ and $K^2>2b$. The associated angular frequencies are
$$\cases{\omega_I= {1 \over (I+ \sqrt {K^2-2b})^3}\cr
\omega_K={K \over \sqrt {K^2-2b}(I+\sqrt {K^2-2b})^3}.\cr}$$
With the help of the action-angle variables, the unperturbed Hamiltonian
${\cal H}_0$ defined
by relation (2) can then be written as
$$H_0(I,K)=-{1 \over 2(I+ \sqrt { K^2-2b})^2},$$
so the unperturbed equations of motion take the form
$$\left\{ \begin{array}{l}
  \.I ={\partial H_0 \over \partial \phi} = 0 \smallskip \\
  \.K= {\partial H_0 \over \partial \th}=0  \smallskip \\ 
 \.\phi= -{\partial H_0 \over \partial I}= \omega_I \smallskip \\ 
 \.\th= -{\partial H_0 \over \partial K}= \omega_K. \end{array}
 \right. \en(16)$$
Recall that the eigenvalues $\lambda$ of the monodromy operator of a {\it
T}-periodic solution are called {\it characteristic multipliers} and that
the numbers 
$\alpha$ defined  by $\lambda= exp(\alpha T)$ are called {\it characteristic
exponents}
\cite{16}, \cite{18}.  We would like to show the nonexistence of other
analytic integrals, 
independent of the Hamiltonian $H_0$. Our main theoretical tool in proving the 
nonintegrability of the perturbed system is the following result, proved in
\cite{11}.  




\begin {theorem} Suppose that the two degree of freedom Hamiltonian system
with Hamiltonian function $H$ has two first integrals $H$ and $F$ that are
independent along a periodic solution. Then four characteristic exponents
vanish.
\end{theorem}

One exponent vanishes because the system is autonomus. The second vanishes
due to 
the presence of the first integral $H$. If the remaining exponents are
different 
from zero, the periodic solution is called {\it nondegenerate}. It is well
known that 
nondegenerate periodic orbits are isolated \cite{12}.

To establish the existence of a large number of isolated periodic orbits,
suppose that for $I=I^0$ and $K=K^0$, the frequencies $\omega_I$ and
$\omega_K$ of the unperturbed problem are commensurable and that $\omega_I \neq
0$. Then the perturbing function $W(I,K,\omega_I t, \omega_K t + \lambda)$,
defined
by $\cal W$ given in (2) and then transformed with the help of the
action-angle variables, 
is periodic in $t$ and has period $T$. Consider its average
$${\bar W}(I^0,K^0,\lambda)={1\over
T}\int_{0}^{T}W(I,K,\omega_It,\omega_Kt+\lambda)dt.$$
Then the following theorem, also due to Poincar\'e \cite{11}, establishes
the existence of 
isolated periodic orbits.    

\begin{theorem} Assume that the following conditions are satisfied:

(1)The Hamiltonian is nondegenerate at the point $I=I^0, K=K^0$,   

(2) for some $\lambda= \lambda^{*}$ the derivative $\partial \overline{W}/
\partial \lambda=0$, but $\partial^2 \overline W/ \partial \lambda^2 \neq 0$.

\noindent Then, for small $\ep \neq 0$, the perturbed Hamiltonian system has
a {\it
T}-periodic solution that depends analytically on the parameter
$\ep$ and for $\ep=0$ coincides with the periodic solution
$$I=I^0,\ \ K=K^0,\ \ \phi= \omega_I T,\ \ \theta=\omega_K t +\lambda^{*}$$
of the unperturbed system. The two charateristic exponents $\pm \alpha$ of
this solution admit a convergent series expansion  in power series of $\sqrt
{\ep}$,
$$ \alpha =\alpha_1 \sqrt{\ep}+ \alpha_2 \ep + \alpha_3 \ep \sqrt{ \ep  }+
\dots, \en(17)$$
where 
$$\omega_I^2\alpha_1^2={\partial^2 \overline {W} (\lambda^{*}) \over
\partial \lambda^2} \left( \omega_I^2 { \partial ^2 H_0 \over \partial K^2}-
2\omega_I \omega_K { \partial^2 H_0 \over \partial I \partial K} +
\omega_K^2 { \partial^2 H_0 \over \partial I^2} \right). \en(18).$$ 
\end{theorem}

problem. 
Using Theorem 2, one can show the dependence of the functions $H_0$ and $F_0$ on
the set of unperturbed tori $I=I_0, \, K=K_0$ that satisfy the conditions of
Theorem 3 and the relation
$$\left( \omega_I^2 { \partial ^2 H_0 \over \partial K^2}- 2\omega_I
\omega_K { \partial^2 H_0 \over \partial I \partial K} + \omega_K^2 {
\partial^2 H_0 \over \partial I^2} \right) \neq 0. \en(19)$$
So the following result holds.

\begin{theorem}
Under the hypotheses of Theorem 3 and relation (19),  the Hamiltonian 
system (16) does not admit a first integral $F$ independent of $H$ that can
be written 
as a formal power series of the form $\sum_{s \geq 0} F_s(I,K,\phi,\theta)
\epsilon^s $ 
with analytic coefficients $F_s$. 
\end{theorem}

\section*{\large\bf V. Fourier series of the perturbation}

\noindent In order to apply these results to our perturbed problem, we first
need to write
the perturbation function as a Fourier series. For this let us first notice
that $r$ can 
be expressed in parametric form with respect to time. Following the same
procedure used 
in \cite{19} for the Kepler problem, we can write
$$t=\sqrt{{1 \over 2 |h|}}  \int {r dr \over  \sqrt{ -r^2 +  {r \over |h|}-
{ K^2 -2b \over 2 |h|}} }.$$
With the substitutions
$$A={1 \over 2|h|} \qq {\rm and}  \qq A^2e^2= \Bigg({1 \over 2|h|}\Bigg)^2
-{K^2 -2b
\over 2|h|},$$
the above relation takes the form
$$t=\sqrt {A} \int { {r dr \over \sqrt{ A^2e^2-(r-A)^2}}}.$$
With the  change of variables $ r-A=-Ae \cos\eta$, the integral assumes the
simpler expression
$$t= A^{3 \over 2} \int (1-e \cos \eta ) d \eta= A^{3 \over 2}(\eta -e \sin
\eta ) + {\rm constant}.$$
If we choose the time in such a way that the addition constant is zero, then
$$r= A(1- e \cos \eta) \qq {\rm and}  \qq t= A^{3 \over 2}(\eta -e \sin \eta).$$
We can now write the constants $A$ and $e$ in terms of the action variables as
$$A={1 \over 2|h|}= (I+ \sqrt {K^2-2b})^2 \qq {\rm and} \qq
e^2={I(I+2 \sqrt {K^2-2b}) \over (I+ \sqrt {K^2 -2b})^2}.$$
Recall that the Fourier series expansion of a function $F$ has the form 
$$F(\eta)= \sum \limits_{l= -\infty}^{+ \infty} A_l e^{il \eta}, \ {\rm
where} \ \ A_l={1
\over 2 \pi} \int_{0}^{2 \pi} F( \eta) e^{-il \eta} d \eta,$$ 
and that that the Bessel function $J_m(z)$ is given by
$$J_m(z)= {1 \over 2 \pi} \int_{0}^{2 \pi} \cos (mu-z \sin
u)du=(-1)^mJ_{-m}(z).$$
It is now easy to show \cite{19} that
$${A \over r}= 1+ 2 \sum_{l=1}^{ \infty } J_l(le) \cos l\eta,$$
where $\eta =\omega_I t$. Therefore
$${A^2 \over r^2}=1 + 4 \sum_{l=1}^{ \infty } J_l(le) \cos l\eta + 4
\sum_{l,m=1}^{
\infty} J_l(le)J_m(me) \cos l \eta \cos m \eta.$$
Using some trigonometry we obtain that 
$$\matrix{\vspace{0.3cm}{A^2 \over r^2}=1 + 4 \sum_{l=1}^{ \infty } J_l(le)
\cos l\eta (l+m) +\cr
2 \sum_{l,m=1}^{\infty} J_l(le)J_m(me) [ \cos \eta (l+m) + \cos \eta(l-m)
].\cr}$$
More computations lead us to the relation 
$${A^2 \over r^2}=1+ \sum_{l=1}^{\infty} (J_l(le))^2  + 4 \sum_{l=1}^{\infty}
\Big(J_l(le)+{A_l \over 2} +B_l \Big) \cos \eta l -2A_1 \cos \eta,$$
where
$$A_l= \sum_{\stackrel{\alpha,\beta}{{\scriptscriptstyle \alpha+\beta=l}}}
J_{\alpha}(\alpha e)J_{\beta}(\beta e)\qq {\rm and} \qq
B_l=\sum_{\stackrel{\alpha,\beta}{{\scriptscriptstyle \alpha-\beta=l}}}
J_{\alpha}(\alpha e)J_{\beta}(\beta e).$$  
We can write the the perturbation function $W$ as a series of the form
$$W(I,K,\omega_It,\omega_K t)=\Big (D+\sum_{l=1}^{\infty}C_l\cos l
\omega_I t\Big)\cos^2\omega_K t,$$
where 
$$C_l={4b\over A^2}\Big[J_l(le)\Big({A \over b}+1\Big)+{A_l \over 2}+
B_l \Big] \ \ {\rm for}\ \  l \neq 1,$$
$$C_1={4b\over A^2}\Big[J_l(le)\Big({A \over b}+1\Big)+{A_l \over 2}+
B_l\Big]-2A_1,$$
and 
$$D={b \over A^2}\Big(1+ \sum_{l=1}^{\infty}(J_l(le))^2 \Big).$$
Using some trigonometry, the above relation becomes
$$W(I,K,\omega_It,\omega_K t)=D +{1 \over 2}
\sum_{\stackrel{l=0}{\scriptscriptstyle m=\pm
1,0}}^{\infty} C_{l,m} \cos \big(l\, \omega_I+ 2 m \,\omega_K \big)\,t,$$
where $C_{l,m}= C_l$ for all $l,m \neq 0$, $C_{0,1}=D$, and $C_{0,-1}=0$.

\section*{\large\bf VI. Periodic orbits and nonintegrability}

\noindent We can now apply the method of the previous section to the
anisotropic 
Manev problem. As we are going to show, all the assumptions of Theorem 3 are
satisfied, 
so for weak anisotropy the existence of isolated periodic orbits follows.
This leads to
the following result.

\begin{theorem} For small $\epsilon \neq 0$, the weakly anisotropic Manev
problem has a 
{\it T}-periodic solution that depends analytically on the parameter 
$\epsilon$. For $\epsilon=0$ this solution coincides with the periodic solution
$$I=I^0,\q K=K^0, \q \phi=\omega_IT, \q \theta=\omega_K t+\lambda^*$$
of the unperturbed system, where $ \lambda^*$ can take the values 
$0,{\pi \over 2}, \pi,{3\pi \over 2}$. Moreover, the two charateristic 
exponents $\pm \alpha$ of the solution admit a convergent series expansion 
in $\sqrt{\epsilon}$, given by (17), where $\alpha_1$ is defined by (18). 
\end{theorem}
{\bf Proof.} 
First we will show that the assumptions of Theorem 3 are satisfied. 
It is easy to see that the unperturbed Hamiltonian is non-degenerate. Indeed,
$$det \left( \begin{array}{lr} {\partial^2 H_0 \over \partial I^2} &
{\partial^2 H_0 \over \partial I \partial K } \medskip \\ {\partial^2 H_0
\over \partial K \partial I} & {\partial^2 H_0 \over \partial K^2}
\end{array} \right)={b \over (K^2-2b)^{3/2} (I+ \sqrt{K^2-2b})^3}.$$
We are left with computing the average $\bar W$ of the perturbation function 
$W$ for values of the action variables whose frequencies of the unperturbed
problem are commensurable. Let $\omega_K={p\over q}\omega_I$. Then we can write
the perturbation as
$$D +{1 \over 2} \sum_{\stackrel{l=0}{\scriptscriptstyle m= \pm
1,0}}^{\infty} C_{l,m} \cos \left [\left(l+2 m {p \over q} \right)\omega_I
\, t+2m \lambda \right].$$
It is clear that the only $\lambda$-dependent terms that survive after
averaging 
are the resonant ones, i.e., the terms with the property that $l+2m{p\over
q}=0$.
It is easy to see that we have to consider only the terms with $m=\pm1$.
This implies $q= \pm 1, \pm 2$ and then $l= \pm 2p, \pm p$. These terms
are of the form
$$C_{\pm 2p,\pm 1} \cos \left(2 \lambda \right) \q {\rm and}\q C_{\pm
p,\pm 1}\cos \left(2 \lambda \right),$$
so 
$${\partial\overline{W}\over\partial\lambda}=G\sin 2\lambda
\q{\rm and}\q{\partial^2 \overline{W}\over\partial\lambda^2}=2G\cos 2\lambda,$$
where 
$$G=2 \left( C_{-2p,1} + C_{2p,1}+ C_{-p,1}+C_{p,1} \right).$$
This means that for $\lambda^*=0,\, {\pi \over 2}, \pi,{3 \pi \over 2}$, the
second condition of Theorem 3 is satisfied if $G$ is not identically zero.

We will now show that, in general, $G$ cannot be identically zero. Indeed,
notice first that $G$ depends on $A$ and is analytic in $A$. But $G$ cannot be
identically zero beyond a discrete set of values of $A$ unless the Bessel
function is 
itself identically zero. However, the Bessel function has only a discrete
set of zeroes. 
Therefore, generically, $G$ cannot be identically zero, so the existence of
isolated 
periodic orbits follows. This completes the proof. 

\medskip

For the anisotropic Manev problem, we are now able to prove the nonexistence
of first 
integrals that are independent of $H$ and analytic in the parameter $\ep$.
In more 
formal terms, the following result holds.

\begin{theorem} For small values of $\ep$, the weakly anisotropic Manev
problem does not 
admit a formal first integral $F$ independent of $H$ that can be written as
a power 
series $\sum_{s \geq 0} F_s(I,K, \phi, \theta) \ep ^s$ whose coefficients
$F_s$ are
analytic functions.
\end{theorem}
{\bf Proof.} According to Theorem~4, the only thing we need to prove is that 
inequality (19) is satisfied. Then using Theorem 5 the result follows. In our 
case inequality (19) can be written as
$$\omega_I^3-2\omega_K^2\omega_I+{\omega_K^4\over\omega_I}+b{\omega_I^{11/3} 
\over \omega_K} \neq 0.$$
If we substitute $\omega_K={p\over q}\omega_I$ (where $\omega_I\neq 0$ and
$p\neq 0$) 
we obtain that (19) is always verified except for the solutions of the equation
$$\omega_I^3 \left({p \over q}\right)^5 -2\omega_I^3\left({p \over q}\right)^3+
\omega_I^3\left({p \over q}\right)+b\omega_I^{8/3} = 0.$$
But this equation has at most five distinct solutions, which do not affect
the outcome 
since, if eliminating them, the remainig set is still dense in the real line. 
This completes the proof.


\section*{\large\bf VII. Geodesic deviation}

\noindent In this and the following section we will show that if the
particles do not collide 
but stay close enough together, the motion is not chaotic. This is somewhat
surprising 
since one expects that, as in celestial mechanics or atomic physics, chaos
appears 
because of near-collision approaches. Of course, our result does not exclude
the possibility 
that chaos is determined by only brief passages close to collision.

Our proof is based on a local criterion due to Szcz\c esny and Dobrowolski
\cite{13}.
The idea is to rewrite the equations of motion in terms of the geodesic
equation 
with respect to the Jacobi metric and to measure the local tendency of the
geodesics to diverge or converge. If they diverge, chaos may appear, but if
they 
converge, chaos is impossible.

Let us first outline the ideas that lead to the above mentioned criterion.
In general, 
for a Hamiltonian function
$${\cal H}({\bf q},{\bf p})={1\over 2}g^{ij}({\bf q}){\bf p}_i{\bf
p}_j+U({\bf q}),$$
where $g^{ij}$ are the components of the covariant tensor corresponding
to the Riemannian metric $g_{ij}d{\bf q}^i\otimes{\bf q}^j$, with ${\bf
p}_i=g_{ij}{\bf q}^i$,
the equations of motion can be written as the Euler-Lagrange system
$${\ddot{\bf q}}^i+\Gamma^i_{\ kl}{\dot{\bf q}}^k{\dot{\bf
q}}^l=-g^{ik}\partial_kU({\bf q}),
\en(20)$$
in which $\Gamma^i_{\ kl}$ is the Levi-Civita connection for the metric
$g_{ij}$. The energy relation ${\cal H}({\bf q},{\bf p})=h$ foliates the
phase space in
codimension-one manifolds. Let us fix an $h$ and define an open set ${\bf
D}_h$ of the
configuration space as ${\bf D}_h=\{{\bf q}\ \! | \ \! U({\bf q})<h\}$. It
can be shown
that the orbits of (20) are orthogonal to the boundary of ${\bf D}_h$. Let
us define the
Jacobi metric ${\hat g}^{ij}=2(h+U)g^{ij}$ and the parameter $s$ such
${ds\over dt}=
2(h+U)$. It can be shown that on the set ${\bf D}_h$, equations (20) are
equivalent 
to the geodesic equations 
$${d^2{\bf q}^i\over ds^2}+{\hat\Gamma}^i_{\ jk}{d{\bf q}^j\over ds}{d{\bf
q}^k\over ds}=0,\en(21)$$
where ${\hat\Gamma}^i_{\ jk}$ is the Christoffel symbol with respect to the
metric ${\hat g}^{ij}$. Note that ${\hat g}_{ij}({\bf q}(s)){d{\bf q}^i\over ds}
{d{\bf q}^j\over ds}=1$. Equations (21) can be further written as
$${\hat\nabla}_{\bf u}{\bf u}={\bf 0},$$
where $\bf u$ is the tangent vector to the geodesic and ${\hat\nabla}$ is
the covariant derivative with respect to the Jacobi metric. Aiming to
understand the 
relative motion of geodesics in a domain inside ${\bf D}_h$, we take a curve
transversal
to the geodesics, then take a copy of it by transporting each point of the
initial curve 
along the geodesic, and measure in this way the local relative deviation of the
geodesics. It can be proved that, in classical form, the measure of the
deviation is given
by the solution of the Jacobi equation \cite{20}
$${d^2x\over ds^2}=-{\cal K}(s)x,\en(22)$$
where $x$ is the variable that measures the deviation and $\cal K$ is the Gauss
curvature
at $x(s)$.
If ${\cal K}>0$ the geodesics approach each other, whereas if ${\cal K}<0$
they diverge.
For 2-degree
of freedom systems the curvature is given by the formula \cite{13}
$${\cal K}={g^{ij}(\partial_iV)(\partial_jV)+(h-U)\Delta U\over
4(h-U)^3},\en(23)$$
where $\Delta = \partial_1^2+\partial_2^2$ is the Laplacian operator. In
\cite{13} the following result is proved.
\begin{theorem}
If $\cal K$ given by formula (23) is negative, the geodesics diverge and
chaos may appear; if $\cal K$ is positive, the geodesics converge and
chaos cannot take place.\label{SD}
\end{theorem}

We will further use this result to investigate the possibility of
encountering chaotic
motion in the neighborhood of collisions.

\section*{\large\bf VIII. Absence of chaos near collision}

\noindent In order to apply Theorem 1 to the equations of motion of the
anisotropic
Manev problem, let us first show that the hypothesis that leads to the geodesic 
deviation equation (22) is satisfied. For this we need to make sure that the
geodesics 
do not reach the boundary of the set ${\bf D}_h$, in other words we must
isolate a set 
of solutions with this property. This will be achieved while proving the
following
result, which shows that collisionless solutions that keep the particles
close together 
cannot be chaotic.
\begin{theorem} There exists a positive-measure set of orbits that are free
of collisions 
and for which the distance between particles does not exceed a certain
uniform bound. 
Within this set, chaotic motion is ruled out.
\end{theorem}
{\bf Proof.} In the case of the anisotropic Manev problem, the expression of
the kinetic 
energy shows that the metric is flat and that $g_{ij}=1$ for $i=j$ and
$g_{ij}=0$ for 
$i\ne j$. Therefore the curvature $\cal K$ given by formula (23) takes the form
$${\cal K}={|\nabla U|^2+(h-U)\Delta U\over 4(h-U)^3},\en(24)$$
where $U({\bf q})=-{1\over \sqrt{\mu q_1^2+q_2^2}}-{b\over{\mu q_1^2+q_2^2}}$.
Let us now see that there exists a set of solutions, whose projection in the
configuration space lies inside ${\bf D}_h$. These will form an open
set of uniformly bounded and collisionless solutions.   
 
For this consider the analytic diffeomorphic transformation of the dependent
variables,
$$\cases{r=(q_1^2+q_2^2)^{1\over 2}\cr
         \varphi=\arctan(q_2/q_1)\cr
         v=\dot r=q_1p_1+q_2p_2\cr
         u=r^2\dot\varphi=q_1p_2-q_2p_1,\cr}$$
and the analytic diffeomorphic transformation of the independent variable
$$d\tau=r^{-1}dt.\eqno(25)$$
Notice that $r$ and $\varphi$ are polar coordinates, whereas $v$ and $u$
represent the 
rescaled radial and tangential components of the velocity. In these new
variables, the 
equations of motion take the form 
$$\cases{r'=v\cr
       v'=2hr+(\mu\cos^2\varphi+\sin^2\varphi)^{-1/2}\cr
       \varphi'=r^{-1}u\cr
       u'={\mu-1\over 2}r^{-1}[({\mu\cos^2\varphi+\sin^2\varphi})^{-3/2}+
2b({\mu\cos^2\varphi+\sin^2\varphi})^{-2}]\sin{2\varphi}\cr}\eqno(26)$$
and the energy relation ${\cal H}({\bf q},{\bf p})=h$ becomes 
$$u^2+v^2-2r({\mu\cos^2\varphi+\sin^2\varphi})^{-1/2}-2b({\mu\cos^2\varphi+\
sin^2
\varphi})^{-1}=2hr^2.\eqno(27)$$
The new dependent variables  $(r,v,\varphi,u)\in (0,\infty)\times\Re\times
S^1\times\Re$ are functions of the fictitious time $\tau$, so the prime
denotes from now
on differentiation with respect to this new independent variable. 

Let us first notice that there exist two constants, $m_1$ and $m_2$ with
$0<m_1<m_2<\infty$, 
such that along any solution $(r,v,\varphi,u)$ of system (26), the function
of $\tau$
given by $\mu\cos^2\varphi+\sin^2\varphi$ satisfies the relations
$$m_1<\mu\cos^2\varphi+\sin^2\varphi<m_2,$$
for all $\tau$ for which $\varphi$ is defined. Consequently there exist two
constants, $M_1$ and $M_2$, with $0<M_1<M_2<\infty$, such that
$$M_1<({\mu\cos^2\varphi+\sin^2\varphi})^{-1/2}<M_2\eqno(28)$$
for all $\tau$. From the first two equations of system (26), we obtain the
nonhomogeneous second-order equation
$$r''-2hr=(\mu\cos^2\varphi+\sin^2\varphi)^{-1/2},\eqno(29)$$
which represents a forced harmonic oscillator. We will further ignore the
last two equations of system (26) and retain only the information
that the forcing function $(\mu\cos^2\varphi+\sin^2\varphi)^{-1/2}$ is
uniformly bounded from above and away from zero from below, as relations
(28) indicate. 

We will further restrict our analysis to the invariant manifold of solutions
of negative energy, $h<0$, which obviously exists according to relation (27).
Solving the homogeneous equation $r''-2rh=0$ and then applying to equation (29) 
the method of variation of parameters, we obtain the general solution of
equation
(29) in the form
$$\matrix{\vspace{0.2cm}
r(\tau)&=\Big(c_1+{1\over\sqrt{-2h}}\int_0^{\tau}{\cos(\sqrt{-2h}\sigma)\over
\sqrt{\mu\cos^2\varphi(\sigma)+\sin^2\varphi(\sigma)}}d\sigma\Big)\sin(\sqrt{-
2h}\tau)\cr
{}&\ +\Big(c_2-{1\over\sqrt{-2h}}\int_0^{\tau}{\sin(\sqrt{-2h}\sigma)\over
\sqrt{\mu\cos^2\varphi(\sigma)+\sin^2\varphi(\sigma)}}d\sigma\Big)\cos(\sqrt{-
2h}\tau).}$$
If, for every solution of constants $c_1$ and $c_2$, we apply to each of the
above
integrals the intermediate-value theorem and then use some trigonometry we
obtain 
$$\matrix{\vspace{0.2cm}
r(\tau)=(c_1^2+c_2^2)^{1\over 2}\cos(\omega\tau-\omega_0)-\cr
{\sin^2(\sqrt{-2h}\tau)\over2h\sqrt{\mu\cos^2\varphi(\tau_1)+\sin^2\varphi(\
tau_1)}}
-{\cos^2(\sqrt{-2h}\tau)\over2h\sqrt{\mu\cos^2\varphi(\tau_2)+\sin^2\varphi(
\tau_2)}},
\cr}\eqno(30)$$
where $\tau_1$ and $\tau_2$ belong to the interval $(0,\tau)$,
$\omega=\sqrt{-2h}$ and 
$\omega_0=\arctan{c_2\over c_1}$. Relations (28) and (30) allow us to draw
the conclusion 
that for any solution $(r,v,\varphi,u)$ of system (26), the inequalities
$$\matrix{
\sqrt{c_1^2+c_2^2}\cos(\omega\tau-\omega_0)-{M_1\over 2h}<
r(\tau)<\sqrt{c_1^2+c_2^2}\cos(\omega\tau-\omega_0)-{M_2\over 2h}}\eqno(31)$$
take place for all $\tau$ for which the solution is defined.

Let us now fix some $h_0<0$, as close to 0 as we like, and define the set
$\Lambda(h_0)$ of solutions formed by the union of all sets $\Lambda_h$, for
$h$ in the
interval $(-\infty,h_0)$, where each set $\Lambda_h$ contains all the
solutions given 
by (30) that satisfy the inequality
$$c_1^2+c_2^2<{M_1^2\over 4h^2}.$$
Obviously, this is an open, nonempty, and connected invariant manifold of
system (26). From (31) it follows that, for every solution of this manifold, $r$
is positive and bounded, therefore the orbits are collisionless and bounded.
Let us
denote by ${\bar\Lambda}(h_0)$ the projection of $\Lambda(h_0)$ into the
configuration
space. For $\mu$ fixed, suitable choices of $c_1, c_2$, and $h$ provide
solutions with 
$q_1$ and $q_2$ small enough such that $U({\bf q})<h_0$. Moreover these
choices can be made 
such that $r$ is as close to $0$ as we like but still positive. Obviously,
the set of the
corresponding solutions has positive measure. Let us denote by $\Gamma(h_0)$
the corresponding 
projection of the set of such solutions in the configuration space.
Obviously $\Gamma(h_0)$ is 
contained in ${\bar\Lambda}(h_0)$ as well as in ${\bf D}_{h_0}$. This shows
that the hypothesis 
leading to the criterion of the previous section is satisfied. A
straightforward computation 
yields 
$$\matrix{{\cal K}=-{3(h-U)(\mu^2q_1^2+q_2^2)\over(\mu q_1^2+q_2^2)^{5/2}}-
{[1-8b(h-U)](\mu^2q_1^2+q_2^2)\over(\mu q_1^2+q_2^2)^3}+
{4b^2(\mu^2q_1^2+q_2^2)\over(\mu q_1^2+q_2^2)^4}+{4b(\mu^2q_1^2+q_2^2)
\over(\mu q_1^2+q_2^2)^7}}.$$
The first term is always negative and the second can also be negative,
depending on
the fixed value of $b$. But since the last two terms are positive and the
powers at
their denominators exceed those of the denominators of the first two terms, for
sufficiently small $q_1$ and $q_2$, the curvature $\cal K$ is positive. We
can always 
define $\Gamma(h_0)$ such that the configurations corresponding to the
solutions with 
corresponding $q_1$ and $q_2$ are contained in $\Gamma(h_0)$. Therefore,
according to 
Theorem~7, the set of collisionless solutions for which the bodies stay
close enough 
together, does not exhibit chaotic behavior. This completes the proof.

\bigskip

\noindent{\bf Acknowledgements.} Florin Diacu was supported in part by the
Pacific Institute for the Mathematical Sciences and by the NSERC Grant
OGP0122045.
Manuele Santoprete was sponsored by a University of Victoria Fellowship.

\vskip0.8cm


\end{document}